\documentclass{article}
\usepackage{ijcai24}

\usepackage{times}
\usepackage{soul}
\usepackage{url}
\usepackage[hidelinks]{hyperref}
\usepackage[utf8]{inputenc}
\usepackage[small]{caption}

\usepackage{graphicx}
\usepackage{amsmath}
\usepackage{amsthm}
\usepackage{booktabs}
\usepackage{algorithm}
\usepackage{algorithmic}
\usepackage[switch]{lineno}

\usepackage{natbib}
\usepackage{array}
\usepackage{amsmath}
\usepackage{subcaption}
\usepackage{makecell}
\usepackage{multirow}
\usepackage{amssymb}  
\usepackage{bm}

\title{CoheDancers: Enhancing Interactive Group Dance Generation \\
through Music-Driven Coherence Decomposition}

\author{
Kaixing Yang\textsuperscript{\rm 1,*},
Xulong Tang\textsuperscript{\rm 2,*},
Haoyu Wu\textsuperscript{\rm 1,*},
Qinliang Xue\textsuperscript{\rm 1},\\
Biao Qin\textsuperscript{\rm 1},
Hongyan Liu\textsuperscript{\rm 3,$\dagger$},
Zhaoxin Fan\textsuperscript{\rm 1,$\dagger$}
\affiliations
\textsuperscript{\rm 1}Renmin University of China, Beijing, China\\
\textsuperscript{\rm 2}Malou Tech Inc, Plano, Texas, USA\\
\textsuperscript{\rm 3}Tsinghua University, Beijing, China\\
\emails
yangkaixing@ruc.edu.cn, Xulong.Tang@maloutech.com, wuhaoyu556@ruc.edu.cn,\\
xueql@ruc.edu.cn,
liuhy@sem.tsinghua.edu.cn, qinbiao@ruc.edu.cn, fanzhaoxin@ruc.edu.cn
}

\begin{document}

\maketitle

\footnotetext[1]{* Indicates equal contribution.}
\footnotetext[2]{$\dagger$ Corresponding authors.}
\begin{abstract}
Dance generation is crucial and challenging, particularly in domains like dance performance and virtual gaming. In the current body of literature, most methodologies focus on Solo Music2Dance. While there are efforts directed towards Group Music2Dance, these often suffer from a lack of \emph{coherence}, resulting in aesthetically poor dance performances. Thus, we introduce \textit{CoheDancers}, a novel framework for Music-Driven Interactive Group Dance Generation. \textit{CoheDancers} aims to enhance group dance generation coherence by decomposing it into three key aspects: synchronization, naturalness, and fluidity. Correspondingly, we develop a Cycle Consistency based Dance Synchronization strategy to foster music-dance correspondences, an Auto-Regressive-based Exposure Bias Correction strategy  to enhance the fluidity of the generated dances, and an Adversarial Training Strategy   to augment the naturalness of the group dance output. Collectively, these strategies enable \textit{CohdeDancers} to produce highly coherent group dances with  superior quality. Furthermore, to establish better benchmarks for Group Music2Dance, we construct the most diverse and comprehensive open-source dataset to date, \textit{I-Dancers}, featuring rich dancer interactions, and create comprehensive evaluation metrics. Experimental evaluations on \textit{I-Dancers} and other extant datasets substantiate that \textit{CoheDancers} achieves unprecedented state-of-the-art performance. Code will be released.
\end{abstract}

\section{Introduction}
\par
Music2Dance is an essential yet challenging task that merges the deep emotional and aesthetic expressions of dance with music, transforming dance performance and education. This task gains complexity in group dance settings, where synchronization between individual dancers and the entire ensemble is critical. Group Music2Dance tasks, therefore, present more significant challenges and offer richer research opportunities than Solo Music2Dance tasks, emphasizing the task's importance and the intricate dynamics involved.

\par
However, though significant progress has been made in Solo Music2Dance through advanced technologies, like variational autoencoders \cite{siyao2022bailando, siyao2023bailando++}, generative adversarial networks \cite{kim2022brand}, and diffusion models \cite{tseng2023edge,li2024lodge}, Group Music2Dance tasks continue to face substantial challenges. Specifically, existing methods typically extend solo techniques by merely adding a basic interaction layer among dancers \cite{yang2024codancers,le2023music,le2023controllable}. This approach often leads to incoherent sequences, underscoring the need for more sophisticated modeling techniques. Furthermore, the benchmarks for evaluating these systems are also inadequate. Many datasets either feature repetitive movements \cite{le2023music} or lack sufficient data for effective training \cite{siyao2024duolando}, which further hampers the development of models capable of generating coherent group dances.

\begin{figure}[t]
  \centering
  \includegraphics[width=0.49\textwidth]{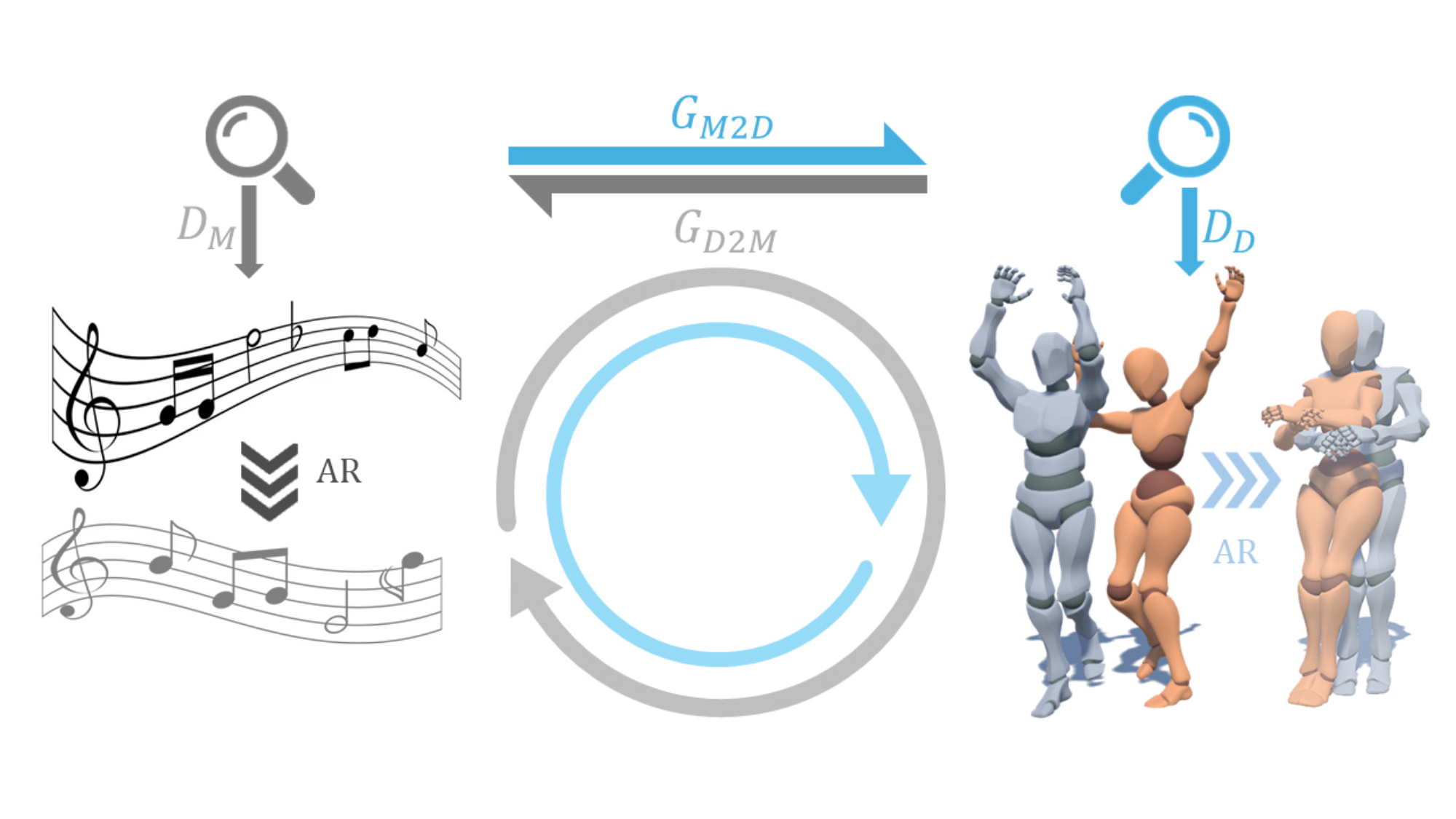}
   \vspace{-0.3in}
  \caption{ CoheDancers synergistically utilizes a Cycle Consistency-based Dance Synchronization strategy, adversarial training, and exposure bias-corrected autoregression to enhance coherence. }
  \vspace{-0.2in}
  \label{fig:intro}
\end{figure}

\par

To tackle the issue, we introduce \textit{CoheDancer}, a novel music-driven group dance generation method in this paper. The intuition behind \textit{CoheDancer} is based on the understanding that the concept of coherence can be dissected into three key components\cite{rallis2018embodied}: synchronization, naturalness, and fluidity.  Consequently, in the stages of model training and network design, it is crucial to specifically tailor designs that focus on these components, thereby boosting the model's overall performance in achieving coherence.

To this end, we develop three innovative strategies within \textit{CoheDancer} to enhance coherence in group dance generation. Firstly, we implement a Cycle Consistency based Dance Synchronization strategy. This approach leverages the inherent interplay between music and dance by constructing generation cycles that accurately reconstruct the original input after each transformation. This ensures synchronization at both group and individual levels through respective dance and music cycles, thus fostering precise dance synchronization. Secondly, an Auto-Regressive-based Exposure Bias Correction strategy improves dance fluidity. While auto-regression simplifies learning and enhances the coherence of generated sequences, it also introduces exposure bias. \textit{CoheDancers} addresses this issue with a specially designed correction strategy that not only enhances fluidity but also ensures the quality of the generated dances. Lastly, an Adversarial Training Strategy  increases dance naturalness. Through this training, the generator progressively refines its ability to emulate authentic dance movements, while the discriminator improves its detection of synthetic movements. This dynamic interaction compels the generator to produce more sophisticated and realistic dance sequences. Collectively, these strategies enable \textit{CoheDancers} to produce highly coherent and superior quality group dances, as demonstrated in Fig. \ref{fig:intro}.

\par
Building on the foundation laid by \textit{CoheDancers}, we also introduce the I-Dancers dataset to further refine benchmarking standards for Group Music2Dance. This extensive open-source dataset, accompanied by an efficient data collection pipeline, contains about 900 top-rated videos from YouTube, covering 3.9 hours across 12 dance genres and showcasing rich dancer interactions. To evaluate the performance effectively, we also develop comprehensive evaluation metrics inspired by music-dance retrieval and dance kinesiology. These metrics assess both local synchronization and global semantic coherence. Our comprehensive quantitative and qualitative experiments conducted on the I-Dancers and AIOZ-GDANCE datasets not only demonstrate the superiority of \textit{CoheDancers} but also validate the effectiveness of the I-Dancers dataset in enhancing the field of music-driven group dance synthesis.

\par
Our contributions can be summarized as: 1) We introduce \textit{CoheDancers}, a novel Group Dance Generation method that strategically decomposes coherence into synchronization, naturalness, and fluidity for enhanced processing. 2) We develop a suite of strategies within \textit{CoheDancers} including a Cycle Consistency based Dance Synchronization strategy to foster music-dance correspondences, an Auto-Regressive-based Exposure Bias Correction strategy to enhance dance fluidity, and an Adversarial Training Strategy  to augment the naturalness of the group dances. 3) We propose the I-Dancers dataset, the most diverse collection to date, accompanied by tailored evaluation metrics designed to refine benchmarking standards for the Group Music2Dance task.

\section{Related Work}
\subsection{Solo Music2Dance Generation}
The field of human motion generation has experienced considerable advancements, particularly within the Music2Dance domain. Researchers have increasingly harnessed musical features—such as those derived from librosa \cite{li2021ai,mcfee2015librosa} and Jukebox \cite{tseng2023edge,dhariwal2020jukebox}—to predict human pose parameters, including SMPL values \cite{loper2023smpl,pavlakos2019expressive,romero2022embodied} and body keypoints \cite{gong2023tm2d}. These developments have catalyzed progress in Solo Music2Dance, with pioneering methods introducing groundbreaking strategies. For instance, the application of Transformer\cite{vaswani2017attention}-structure models by \cite{li2021ai,zhuang2022music2dance,li2022danceformer} to directly predict pose parameters has ignited substantial interest, although the resulting motions often lack coherence and naturalness. The adoption of a GAN framework by \cite{kim2022brand,sun2020deepdance} has notably enhanced the naturalness of the dance motions generated. Moreover, recognizing the repetitive nature of dance movements, studies such as \cite{siyao2022bailando,siyao2023bailando++,zhuang2023gtn} have encoded and quantized meaningful dance components into a choreographic memory, facilitating the generation of aesthetically pleasing dances, albeit at the potential cost of creative freedom. Recent applications of diffusion techniques \cite{tseng2023edge,loper2023smpl,li2023finedance,yao2023dance,zhang2024bidirectional} now enable the direct generation of dance motions from music, circumventing the exposure bias inherent in autoregression but at the expense of computational efficiency and sequence coherence. Despite significant strides in Solo Music2Dance, extending them to group settings introduces additional complexity due to the intricate interactions within group dynamics.

% In this paper, we address the equally vital task of group music generation, examining it through the lens of coherence, thereby contributing to a deeper understanding and enhanced capabilities in this challenging yet fascinating area.

\begin{figure*}[!t]
  \centering
  \includegraphics[width=\textwidth]{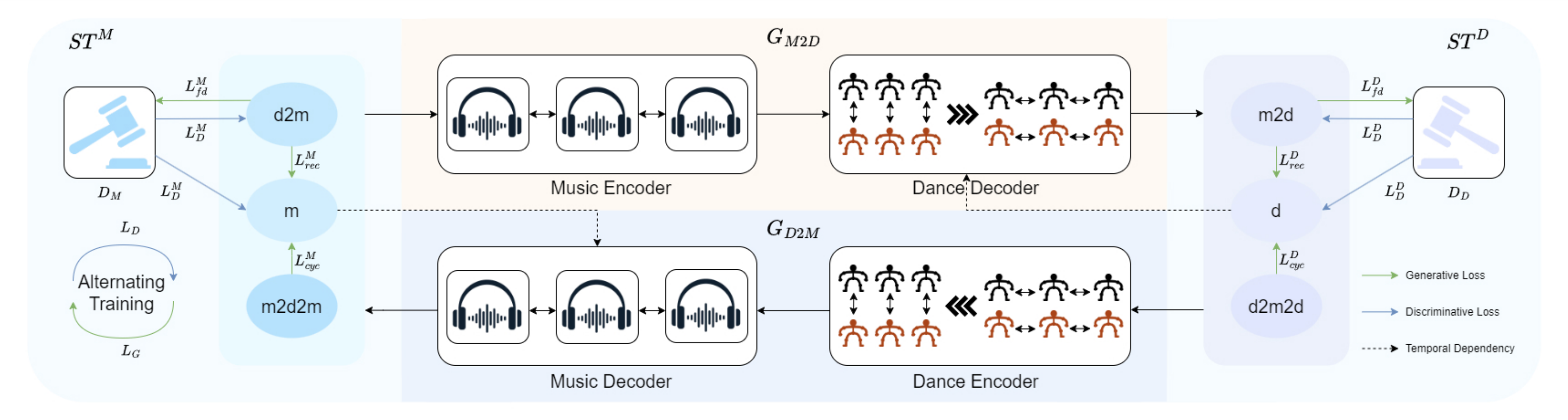}
  \vspace{-0.2in}
  \caption{\textbf{Pipeline Structure}. CoheDancers comprises Music2Dance Generation block ($G_{M2D}$), Dance2Music Generation block ($G_{D2M}$), and Strategy Training block ($ST^{M}, ST^{D}$), working synergistically to enhance dance coherence.}
  \label{fig:overview}
  \vspace{-0.2in}
\end{figure*}

\subsection{Group Music2Dance Generation}
The endeavor to generate group dances has encountered notable challenges at both model and benchmark levels, which has restrained progress within the field. Initially, prevalent approaches in the adaptation of Solo Music2Dance models for group contexts, such as incorporating interaction layers like Transformers or LSTMs, have proven insufficient. Models like FACT \cite{li2021ai} adapted for GDancerR \cite{le2023music,wang2022groupdancer}, Bailando \cite{siyao2022bailando} adapted for CoDancers \cite{yang2024beatdance}, and EDGE \cite{tseng2023edge} adapted for GCD \cite{le2023controllable}, often result in group dances that lack coherence.

Furthermore, the existing benchmarks do not adequately support the complexities of group dance generation. For instance, the dataset provided by \cite{siyao2024duolando} offers high-quality MoCap data but spans only 1.9 hours across ten dance genres, which is insufficient for robust deep learning training. Another dataset by \cite{le2023music} extends to 5 hours of content; however, 80\% of this data showcases repetitive movements among dancers, overlooking the nuanced interactions that are crucial to capturing the essence of group dance. The AIST\cite{tsuchida2019aist} dataset provides only 1.44 hours of group dance data, with limited interaction complexity among dancers. This leads to suboptimal generation outcomes. Moreover, metrics based on Kinetic\cite{onuma2008fmdistance} or Geometric\cite{muller2005efficient} Features have been shown to misalign with artistic assessments of dance, as evidenced by \cite{tseng2023edge}.

Given these significant gaps, there is a substantial opportunity for advancement in models and benchmarks within the Group Music2Dance domain. 

% In this paper, we introduce \textit{CoheDancers}, a model designed to address the critical issue of music-dance coherence by decomposing it into synchronization, naturalness, and fluidity. Additionally, we present the I-Dancers dataset, the most diverse dataset to date, aimed at fair benchmarking of existing methods.

\section{Method}
\subsection{Problem Definition}
Given a sequence of music features ${m_1, m_2, ..., m_T}$ and the initial poses of $N$ dancers ${d^1_0, d^2_0, ..., d^N_0}$, our objective is to synthesize the corresponding group dance sequences ${d^1_1, ..., d^1_T; ...; d^N_1, ..., d^N_T}$, where $d^i_t$ denotes the pose of the $i$-th dancer at time step $t$. We define each music feature as a 438-dimensional vector, extracted via Librosa, and represent each dance motion as a 147-dimensional vector $d = [\tau; \theta]$, where $\tau$ and $\theta$ encapsulate the root translation and 6D rotation pose parameters of the SMPL model, respectively. Furthermore, we synchronize the music feature sequence with the group dance sequences at a temporal granularity of 30 frames per second, ensuring precise alignment between the auditory and movement aspects of the performance.

\subsection{Overview}
We introduce the framework of CoheDancers in separate sections, specifically detailing the Pipeline Structure and  Training Strategy.

In the Pipeline Structure part, as depicted in Fig. \ref{fig:overview}, two main blocks are presented. The Music2Dance Generation block processes the input music feature $m$, generating dance motions $m2d$ through temporal autoregression. Conversely, the Dance2Music Generation block takes dance motions $d$ as input, producing corresponding music features $d2m$:

\begin{align}
\begin{aligned}
m2d &= G_{\mathrm{M2D}}(m) \
d2m &= G_{\mathrm{D2M}}(d)
\end{aligned}
\end{align}

\begin{figure*}[!t]
  \centering
  \includegraphics[width=\textwidth]{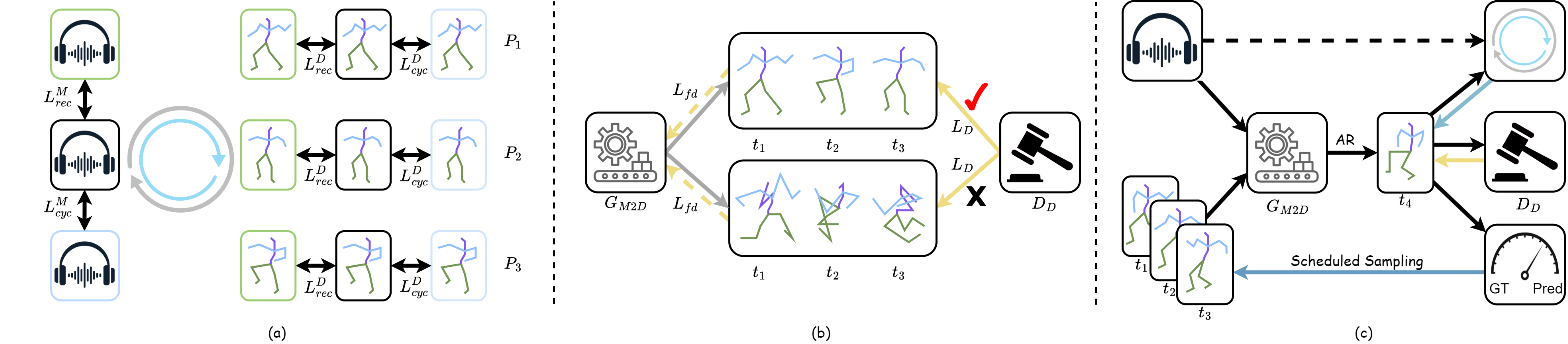}
  \vspace{-0.2in}
  \caption{\textbf{Strategy Training}. Strategy comprises the (a) Cycle Consistency based Dance Synchronization strategy, (b) the Adversarial Training pipeline, and (c) the Auto-Regressive-based Exposure Bias Correction strategy.}
  \label{fig: strategy}
  \vspace{-0.2in}
\end{figure*}

In the Training Strategy part, outlined in Fig. \ref{fig: strategy}, three main strategies are employed: the Cycle Consistency based Dance Synchronization strategy, the Auto-Regressive-based Exposure Bias Correction strategy, and the Adversarial Training pipeline. These strategies leverage the discriminative loss $\mathcal{L}{D}$ and the generative loss $\mathcal{L}{G}$, which are computed based on related music $m$ and dance $d$ information processed by the Pipeline. The losses are then alternately trained to refine the models:

\begin{equation}
\mathcal{L}_{G}, \mathcal{L}_{D} = \mathrm{ST}_(m, d)
\end{equation}

% Next, we introduce the Pipeline Structure part  and Training Strategy Training part in detail.

\subsection{Pipeline Structure}
\subsubsection{Music2Dance Generation Block}
The Music2Dance Generation block is elegantly structured into two components: a Music Encoder and a Dance Decoder.

To adeptly capture trans-temporal nuances, we employ a multi-layer Transformer Encoder to construct the Music Encoder. The attention mechanism is defined as:
 
\begin{equation}
\operatorname{\mathbf{Attention^{me}}} = \operatorname{softmax}\left(\frac{\mathbf{Q_{m}} \mathbf{K_{m}}^{T}}{\sqrt{C}}\right) \mathbf{V_{m}}
\end{equation}
where $\mathbf{Q_{m}, K_{m}, V_{m}}$ are derived from the music features. Post encoding, these features are replicated $N$ times, once for each dancer.

Excelling in group dance necessitates consideration of both temporal and spatial coherence. To address this, we design a multi-layer Dance Decoder, each layer comprising a Spatial Processor and a Temporal Decoder. We utilize a transformer encoder layer without position embedding for the Spatial Processor to capitalize on its position invariance, detailed by the attention mechanism:

\begin{equation}
\label{eq:spatial}
\operatorname{\mathbf{Attention_{S}^{dd}}} = \operatorname{softmax}\left(\frac{\mathbf{Q_{d}} \mathbf{K_{d}}^{T}}{\sqrt{C}}\right) \mathbf{V_{d}}
\end{equation}
where $\mathbf{Q_{d}, K_{d}, V_{d}}$ are derived from dance for spatial interaction. Considering the progressive and dependent nature of dance movements, an autoregressive generation style is adopted to reduce learning complexity and ensure coherence. For temporal decoding, we employ a transformer decoder:

\begin{equation}
\operatorname{\mathbf{Attention^{dd}_{T}}} = \operatorname{softmax}\left(\frac{\mathbf{Q_{d}} \mathbf{K_{m}}^{T} + \mathbf{M}}{\sqrt{C}}\right) \mathbf{V_{m}}
\end{equation}
where $\mathbf{Q_{d}}$ comes from the previous dance sequence, $\mathbf{K_{m}}, \mathbf{V_{m}}$ from the full music sequence, and $\mathbf{M}$ is a lower triangular matrix.

\subsubsection{Dance2Music Generation Block}

The Dance2Music Generation block comprises a Dance Encoder and a Music Decoder, intricately designed to capture and translate spatial-temporal dance data into music.

Specifically, mirroring the structural sophistication of the Dance Decoder, the Dance Encoder is also constructed as a multi-layer entity. Each layer begins with a Spatial Processor, utilizing the attention mechanism detailed as:

\begin{equation}
\operatorname{\mathbf{Attention^{de}_{S}}} = \operatorname{softmax}\left(\frac{\mathbf{Q_{d}} \mathbf{K_{d}}^{T}}{\sqrt{C}}\right) \mathbf{V_{d}}
\end{equation}
where $\mathbf{Q_{d}, K_{d}, V_{d}}$ are meticulously derived from the dance data. Following spatial processing, the Temporal Encoder, a transformer encoder, captures the temporal dynamics:

\begin{equation}
\operatorname{\mathbf{Attention^{de}_{T}}} = \operatorname{softmax}\left(\frac{\mathbf{Q_{d}} \mathbf{K_{d}}^{T}}{\sqrt{C}}\right) \mathbf{V_{d}}
\end{equation}
then, post encoding, averaging operations across different dancers are performed to align the dimensions, ensuring a unified representation.

Given the inherent temporal dependencies and sequential structure of music, an autoregressive approach is  then employed within the Music Decoder to ensure fluidity and harmonic consistency. The attention mechanism for this decoder is defined as:

\begin{equation}
\operatorname{\mathbf{Attention^{md}}} = \operatorname{softmax}\left(\frac{\mathbf{Q_{m}} \mathbf{K_{d}}^{T} + \mathbf{M}}{\sqrt{C}}\right) \mathbf{V_{d}}
\end{equation}
where $\mathbf{Q_{m}}$ is derived from the previous music sequence, $\mathbf{K_{d}}, \mathbf{V_{d}}$ from the full dance sequence, and $\mathbf{M}$ is a lower triangular matrix, facilitating the generation of coherent and contextually rich musical outputs from dance inputs.

\subsection{Training Strategy}
\subsubsection{Cycle Consistency based Dance Synchronization Strategy}
This strategy meticulously accounts for dance synchronization to achieve coherence. Music inherently structures dance by providing a temporal framework, and in turn, dance visually embodies musical expressions. Recognizing the deep-seated relationship between music and dance, the CoheDancers framework constructs dual cycles, aiming to accurately reconstruct the original input after each transformation through the joint training of Music2Dance and Dance2Music tasks. This approach markedly enhances synchronization between music and dance at both group and individual levels. Specifically, the music cycle aligns the original music with the cycled music by incorporating information from group dancers. Simultaneously, the dance cycle aligns the original and cycled dances for each individual dancer, utilizing information derived from the corresponding music.

\par
Initially, CoheDancers generates cycled sequences $d2m2d$ and $m2d2m$ from reconstructed sequences $m2d$ and $d2m$ through the inverse mappings:
\begin{align}
m2d2m &= G_{\mathrm{D2M}}(m2d) \\
d2m2d &= G_{\mathrm{M2D}}(d2m)
\end{align}

\par
Subsequently, an L1 loss is applied between the original sequences $m, d$ and their cycled counterparts $m2d2m, d2m2d$:
\begin{equation}
\mathcal{L}_{\text{cyc}} = \mathbb{E}[|m2d2m-m|] + \mathbb{E}[|d2m2d-d|]
\end{equation}
Additionally, the loss for dance includes not only pose parameters and root translation but also the velocity of root translation to ensure smoothness between frames.

\par
Finally, given the distinct distributions between music and dance, a reconstruction loss $\mathcal{L}_{\text{rec}}$ is utilized to ensure that generated outputs closely match real data. The generator is penalized via L1 Loss for deviations between the reconstructed $d2m, m2d$ and the original $m, d$:
\begin{equation}
\mathcal{L}_{\text{rec}} = \mathbb{E}[|d2m - m|] + \mathbb{E}[|m2d - d|]
\end{equation}
Similarly, the loss for dance also incorporates the velocity of root translation to maintain smoothness.

\subsubsection{Adversarial Training Strategy}
 This module is designed to enhance the naturalness of group dance performances. By incorporating Generative Adversarial Networks (GANs) into generation tasks, the discriminator refines the generator by minimizing the discrepancy between real and generated sequences. This adversarial feedback compels the generator to model complex dependencies, thereby boosting the naturalness of the generated music and dance.

\par
The Music Discriminator $D_{M}$ and Dance Discriminator $D_{D}$ employ multi-layer transformer-based architectures, similar to the Music Encoder and Dance Encoder. A unique aspect of their design is that each encoder layer is followed by a 1D-CNN downsampling layer. After encoding, a cross-temporal and cross-dancer averaging operation is executed to derive a probability score for the entire sequence.

\par
The discriminative loss $\mathcal{L}_{D}$ is constructed using real values $m$ and $d$ as positive samples, and predicted sequences $m2d$ and $d2m$ as negative samples, applying Cross-Entropy (CE) Loss:
\begin{align}
\mathcal{L}_{D} &= \mathbb{E}[\log(1-D_{M}(d2m))] + \mathbb{E}[\log(D_{D}(d))] + \\
&\quad \mathbb{E}[\log(1-D_{D}(m2d))] + \mathbb{E}[\log(D_{M}(m))]
\end{align}

\par
In contrast, the generator strives to produce samples $m2d, d2m$ that deceive the discriminators $D_{M}, D_{D}$:
\begin{equation}
\mathcal{L}_{\text{fd}} = \mathbb{E}[\log(1-D_{M}(d2m))] + \mathbb{E}[\log(1-D_{D}(m2d))]
\end{equation}

\par
Ultimately, the reconstruction, cycle consistency, and fool discriminator losses are amalgamated to formulate the generative loss $\mathcal{L}_{\text{G}}$:
\begin{equation}
\mathcal{L}_{\text{G}}=\mathcal{L}_{\text{rec}} + \mathcal{L}_{\text{cyc}} + \mathcal{L}_{\text{fd}}
\end{equation}

\begin{figure}[!t]
  \centering
  \includegraphics[width=0.48\textwidth]{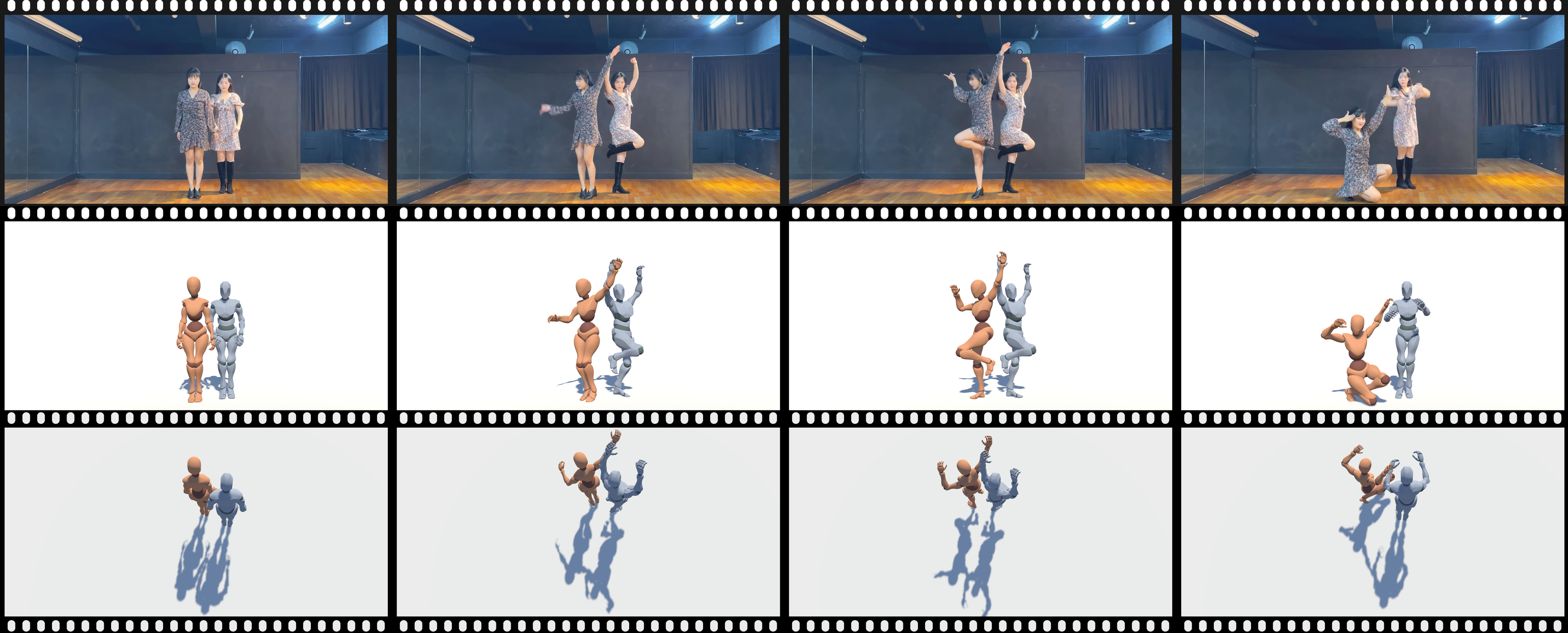}
    \vspace{-0.25in}
  \caption{Dance sequence examples from  I-Dancers dataset.}
  \label{fig:dataset}
  \vspace{-0.2in}
\end{figure}

\subsubsection{Auto-Regressive-based Exposure Bias Correction Strategy}
\par

This strategy is devised to enhance the fluidity of the generated dances. Given the progressive and dependent nature of dance movements, autoregression naturally emerges as a method to improve fluidity. However, the introduction of autoregression introduces the challenge of exposure bias. This phenomenon occurs when a model, trained on actual observed labels, must rely on its own previous outputs during testing. This discrepancy can lead to cumulative errors during inference, adversely impacting the quality of generated outputs. Notably, exposure bias is more pronounced in continuous autoregression tasks like dance generation, compared to discrete tasks such as text generation.

\par
To mitigate this issue, CoheDancers has developed an Exposure Bias Correction strategy during the training phase, thereby enhancing dance fluidity while simultaneously ensuring the quality of generation. Firstly, Scheduled Sampling\cite{bengio2015scheduled} mitigates exposure bias by progressively incorporating the model’s predictions during the training phase, effectively bridging the gap between training and inference scenarios. In our methodology, the integration of predicted sequences increases linearly, facilitating a smooth transition from reliance on ground truth data to self-generated sequences. Secondly, the principle of cycle consistency is employed to ensure that the model can inversely reconstruct the original input following transformations. This inverse reconstruction acts to correct erroneous predictions and prevent the accumulation of errors. Thirdly, the Generative Adversarial Network utilizes a discriminator that is straightforward to train and considers the global sequence context to rectify errors in autoregressive predictions, thereby diminishing the likelihood of error propagation.

\subsubsection{Inference}
\textit{CoheDancer} employs sophisticated training strategies but simplifies inference by using only the $G_{m2d}$ block in Fig. 2, a streamlined transformer-based encoder-decoder architecture. This design achieves real-time generation, producing a 167-second dance sequence in just 47.19 seconds on an NVIDIA RTX A6000 GPU, supported by an Intel Xeon CPU.

\section{Experiment}

\begin{table}[t]
\caption{Comparison with other group dance datasets, including AIST \cite{tsuchida2019aist}, Duolando \cite{siyao2024duolando} and AIOZ-GDANCE \cite{le2023music}. M, H, L represent low, medium, high levels of interactivity respectively. }
\vspace{-0.1in}
\centering
\label{tab: dataset}
% \small
\tiny
\resizebox{0.49\textwidth}{!}{
\begin{tabular}{l|cccc}
\Xhline{1pt}
Dataset & Duration & Interactivity & Genre & Acquisition \\
\Xhline{1pt}
AIST-Group & 1.44h & M & 10 & MoCap \\
Duolando & 1.95h & H & 10 & Pseudo \\
AIOZ-GDANCE(P2) & 4.82h & L & 7 &  Pseudo \\
AIOZ-GDANCE(P3) & 5.13h & L & 7 & Pseudo \\
\hline
I-Dancers & 3.88h & H & 12 & Pseudo \\
\Xhline{1pt}
\end{tabular}
}
\vspace{-0.2in}
\end{table}

\subsection{I-Dancers dataset}
\subsubsection{Data Statistics}
We introduce a substantial dataset, termed I-Dancers, which is rich in dancer interactions and depicted in Fig. \ref{fig:dataset}. This dataset provides an interactive and comprehensive collection of dance data, ideally suited for effective model training. A comparative analysis with related datasets is presented in Tab. \ref{tab: dataset}. I-Dancers encompasses 3.8 hours of whole-body motion and music audio from group dancing, spanning 12 distinct dance genres. Each video in the dataset varies from 20 to 60 seconds in length, meticulously processed at 1080x720 resolution and 30 FPS, ensuring high-quality data for intricate dance motion analysis. See Appendix for details.

\begin{table*}[!t]
\vspace{-0.1in}
\caption{Quantitative evaluation results for Group Music2Dance on AIOZ-GDANCE-P2, AIOZ-GDANCE-P3, and I-Dancers. We compare CoheDancers with Bailando(G) \cite{siyao2022bailando}, CoDancers \cite{yang2024codancers}, FACT(G) \cite{li2021ai}, and GDanceR \cite{le2023music}, and explore the impacts of Dance Synchronization Strategy (DS), Auto-Regressive-based Exposure Bias Correction (EBC) Stragegy and Adversarial Training Strategy (AT).}
\renewcommand\arraystretch{1.2}
\vspace{-0.1in}
\centering
\label{tab: exp}
% \small
\tiny
\resizebox{1.0\textwidth}{!}{
\begin{tabular}{l|l|l|l|l|l|l|l}
\Xhline{1pt}
Datasets  & Methods  & FID $\downarrow$ & M-Dist $\downarrow$ & MM-Dist $\downarrow$ & Div $\uparrow$ & MDA $\uparrow$ & GDA $\uparrow$ \\
\hline
\text{AIOZ-GDANCE-P2} & \textbf{Real Motions} & 00.00 & 00.00 & 18.87 & 19.26 & 0.388 & 0.533 \\
\cline{2-8}
          & Bailando(G) & 100.45 & 16.20 & 20.66 & 17.21 & 0.326 & 0.288 \\
          & CoDancers & 88.14 & 17.18 & 20.82 & \textbf{18.03} & 0.332 & 0.267 \\
          & FACT(G) & 123.59 & 16.58 & 20.50 & 16.43  & \textbf{0.408} & 0.368 \\
          & GDanceR & 104.09 & 15.11 & 20.15 & 16.70 & 0.402 & 0.337 \\
          \cline{2-8}
          & CoheDancers & \textbf{71.18}  & \textbf{13.67} & \textbf{19.40} & 17.54  & 0.405 & \textbf{0.399} \\
          \hline
\text{AIOZ-GDANCE-P3}  & \textbf{Real Motions} & 00.00  & 00.00 & 18.17 & 19.25 & 0.382 & 0.555 \\

\cline{2-8}
          & Bailando(G) & \textbf{67.38}  & 15.65 & 19.87 & 18.19 & 0.337 & 0.281 \\
          & CoDancers & 73.09 & 16.61 & 20.40 & \textbf{18.16} & 0.336 & 0.284 \\
          & FACT(G) & 170.96 & 16.42 & 20.47 & 13.52  & 0.395 & 0.350 \\
          & GDanceR & 121.87 & 15.52 & 19.98 & 15.41 & 0.391 & \textbf{0.469} \\
          \cline{2-8}
          & CoheDancers & 81.63  & \textbf{14.66} & \textbf{19.31} & 17.30 & 0.402 & 0.402 \\
          \hline
\text{I-Dancers}  & \textbf{Real Motions} & 00.00 & 00.00 & 19.74 & 20.03 & 0.389 & 0.574 \\
\cline{2-8}
          & Bailando(G) & 58.31 & 18.50 & 21.35 & 18.72  & 0.321 & 0.273 \\
          & CoDancers & 56.21 & 18.81 & 21.30 & 18.96  & 0.333 & 0.282 \\
          & FACT(G) & 77.19 & 16.95 & 20.57 & 17.82 & \textbf{0.396} & 0.318 \\
          & GDanceR & 76.11 & 16.16 & 20.27 & 17.66 & 0.387 & \textbf{0.334} \\
          \cline{2-8}
          & \textit{- DS} & 70.37 & 16.12 & 20.19 & 17.98  & 0.391 & 0.311 \\
          & \textit{- AT} & 55.92 & 15.71 & 20.20 & 18.40 & 0.392 & 0.308 \\
          & \textit{- EBC} & 130.78  & 15.87 & 20.24 & 15.34  & 0.367  & 0.499 \\
          & CoheDancers & \textbf{37.01} & \textbf{15.50} & \textbf{20.09} & \textbf{19.25} & 0.390 & 0.313 \\
          \cline{2-8}
\Xhline{1pt}
\end{tabular}
}
\vspace{-0.2in}
\end{table*}

\subsubsection{Data Collection}
Given the intricate dynamics of dance and inherent limitations of motion capture technologies, we have established an efficient data collection pipeline for group dance, comprising four key stages: \textbf{(1) Video Collection:} In light of the significant role of dance in contemporary culture and its high demand in entertainment industry, we selectively source high-rated commercial dance videos from YouTube. To mitigate the issues of depth ambiguity and spatial inaccuracies common in monocular pose estimation, we exclusively utilize fixed-camera setups to enhance the accuracy and stability of data capture. \textbf{(2) Pose Estimation:} Utilizing advancements in pose estimation technology, we employ ScoreHMR \cite{stathopoulos2024score} for pose estimation of group dances. This method is chosen for its robustness in managing the complex scenarios presented by group dynamics. \textbf{(3) Pose Processing:} To ensure continuity in dance movements, which is vital, we first apply exponential smoothing. The root transition is directly smoothed, while pose rotations are converted into 6D representations \cite{zhou2019continuity} to address the issue of angle discontinuities. Additionally, we align predicted ground planes across segments to a uniform height, thereby enhancing the consistency of motion. \textbf{(4) Anomaly Detection:} Recognizing the susceptibility of monocular motion capture methods to anomalies, especially in scenarios involving occlusions, anomaly detection is deemed essential. We utilize Blender\cite{blender2024} for visualization, manually identified and removed anomalous frames, and employ the built-in Bezier curve tool to seamlessly complete the segments.

% create table here

\subsection{Evaluation }
\subsubsection{Global Semantic Metrics}
Leveraging efficiency and lower complexity of retrieval tasks, we employ retrieval models for evaluating generative models in areas like text-to-motion \cite{guo2022tm2t}. These models, trained on real data, capture the artistic nuances of dance and music effectively. Our model, inspired by \cite{yang2024beatdance}, combines an Acoustic-Semantic Music Encoder with a Spatial-Temporal Dance Encoder, utilizing Temporal CLIP loss \cite{liu2023revisiting} for contrastive learning. The extracted features are then used to construct robust evaluation metrics such as $\text{FID}$, $\text{M-Dist}$, and $\text{MM-Dist}$, to assess the quality and diversity of dance generations. See Appendix for details.

\subsubsection{Local Synchronized Metrics}
Local synchronization is essential in dance, involving alignment between dance and music and coherence among dancers, with the beat as a key indicator. Drawing from dance kinesiology, moments when the acceleration of human keypoints is zero are identified as dance beats. Music beats are obtained using Librosa, and beat similarity calculations are detailed in \cite{li2021ai}. Our evaluation metrics include Music Dance Alignment (MDA), which focuses on synchronizing the lead dancer’s beats with the music, and Group Dance Alignment (GDA), which assesses beat coordination within the dance group. See Appendix for details.

\subsection{Comparison}
For the I-Dancers dataset, we allocate 85\% for training and 15\% for testing. In the case of the AIOZ-GDANCE dataset \cite{le2023music}, we specifically select data involving two and three dancers, forming the datasets AIOZ-GDANCE-P2 and AIOZ-GDANCE-P3, which are divided into training and testing sets with a 90\% and 10\% distribution, respectively. In the Group Music2Dance domain, our framework benchmarks against established models such as CoDancers \cite{yang2024codancers} and GDanceR \cite{le2023music}. Given the emerging field of Group Music2Dance, we adapt models originally designed for Solo Music2Dance. To ensure fairness and enhance functionality, we modify their original multi-layer architectures by integrating a Transformer at the end of each layer, facilitating effective information exchange among dancers. This modification leads to the development of FACT(G) and Bailando(G).

As demonstrated in Tab. \ref{tab: exp}, CoheDancers significantly outperforms all baseline methods across a range of evaluative metrics. Notably, superior results in FID and M-Dist indicate exceptional semantic alignment between generated and ground-truth dances, while standout performance in M-Dist and MDA suggests highly consistency between generated dance and input music. Moreover, high scores in Div reflect the creative output of CoheDancers, and excellent results in GDA confirm enhanced synchronization among dancers. The remarkable efficacy of CoheDancers is attributed to its synergistic application of cycle consistency (Cycle), generative adversarial training (GAN), and exposure bias correction auto regression (AR) strategies. These strategies collectively empower the model to produce dances with high coherence and superior quality. The pronounced advancements achieved by CoheDancers across all metrics affirm its status as the current state-of-the-art (SOTA) in the field of Group Music2Dance.

\begin{figure}[!t]
  \centering
  \includegraphics[width=0.48\textwidth]{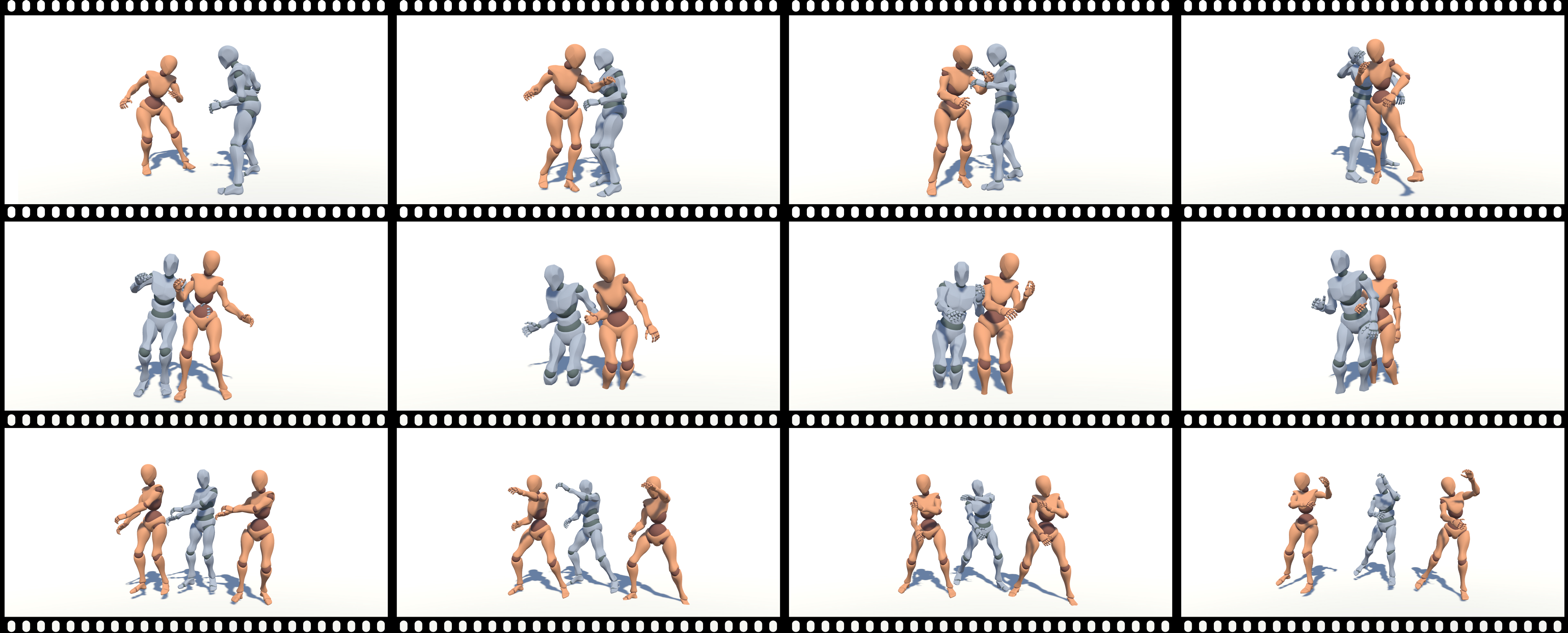}
  \vspace{-0.2in}
  \caption{Dance sequences generated by CoheDancers on I-Dancers, AIOZ-GDANCE-P2 and AIOZ-GDANCE-P3, from top to bottom.}
  \label{fig:experiment}
  \vspace{-0.2in}
\end{figure}

\subsection{Qualitative Analysis}
As illustrated in Fig. \ref{fig:experiment}, CoheDancers demonstrates exceptional visualization across varied datasets, showcasing robust performance in experimental setups with different numbers of dancers. The presented results, derived from models selected based on optimal metrics, partially validate the strategy's effectiveness. In particular, the model trained on the AIOZ-GDANCE dataset achieved visually acceptable outcomes but exhibited significantly lower interactivity compared to the I-Dancer outputs, underscoring its effectiveness. Additionally, the inadequate human-ground contact in the AIOZ-GDANCE dataset led to models where dancers appeared to float or penetrate surfaces, highlighting areas for potential improvement.

\subsection{Ablation Study}
To enhance synchronization between music and dance, we introduced a Cycle Consistency based Dance Synchronization strategy. As shown in Tab. \ref{tab: exp}, beyond improvements in MDA and GDA, Dance Synchronization strategy significantly boosts other metrics, demonstrating its effectiveness. This also suggests that cycle consistency focuses more on aligning global semantics rather than local details.

To make our generated dance more natural, we employ an Adversarial Training pipeline. From Tab. \ref{tab: exp}, the pipeline notably enhances FID, demonstrating its effectiveness. However, the lack of significant improvements in other metrics indicates that the role of our pipeline is more of an enhancement rather than foundation in generation process.

To enhance the fluidity of generated sequences, we introduce an Auto-Regressive-based Exposure Bias Correction strategy. To ablate this, we generate complete dance sequences from only initial dance and music sequences, similar to ActFormer\cite{xu2023actformer}. In group dance generation, autoregression considers prior sequences and music for each dancer as input, while non-autoregression focuses solely on initial frames and music. This causes non-autoregression to struggle with generating complex interactions among dancers, often resulting in more uniform movements. Consequently, while GDA scores improve, other metrics significantly decline, underscoring the effectiveness of the Exposure Bias Correction strategy.

\begin{table}[t]
\caption{User study results on I-Dancers dataset. }
\vspace{-0.1in}
\centering
\label{tab: user study}
% \small
\tiny
\resizebox{0.49\textwidth}{!}{
\begin{tabular}{l|cccc}
\Xhline{1pt}
\textbf{Methods} \& \textbf{Evaluations} & SQ $\uparrow$ & FQ $\uparrow$ & NQ $\uparrow$ \\
\Xhline{1pt}
Ground Truth &  4.16 & 4.54 & 4.63 \\
GDancerR\cite{le2023music} &  3.03 & 3.36 & 3.45 \\
CoDancers\cite{yang2024codancers} & 3.12 & 3.47 & 2.87 \\
CoheDancers (Ours) & \textbf{3.52} & \textbf{3.57} & \textbf{3.94}\\
\hline
Ground Truth &  78\% & 86\%  & 78\%  \\
Generation &  67\%  & 76\%  & 73\%  \\
Mixture &  89\%  & 92\%  & 84\%  \\
\Xhline{1pt}
\end{tabular}
}
\vspace{-0.2in}
\end{table}

\subsection{User Study}
Dance’s inherent subjectivity makes user feedback essential for evaluating generated movements\cite{legrand2009perceiving}, particularly in the Music2Dance task\cite{}. We select 30 music segments (20 seconds each) and generate dance sequences using various models. These sequences are evaluated by 20 undergraduate and graduate students through a blind questionnaire. Three criteria are rated on a 5-point scale: Synchronization Quality (SQ, alignment with rhythm/style), Fluidity Quality (FQ, continuity/aesthetics), and Naturalness Quality (NQ, realism of group dance).

\textbf{Methods}. We compare average scores across methods to evaluate performance. As shown in Table \ref{tab: user study}, CoheDancers achieves SQ = 3.52, FQ = 3.57, and NQ = 3.94 on the I-Dancers dataset, outperforming other methods but remaining below the ground truth. Its strong results across synchronization, fluidity, and naturalness underscore the effectiveness of decomposing dance coherence into these key dimensions.

\textbf{Evaluations}. To validate metric reliability, we assess metric-user alignment using 100 dance pairs under three conditions\cite{li2024mvbench}: (1) Ground Truth (all pairs from real dances), (2) Generation (all pairs from generated dances), and (3) Mixture (half real, half generated). As shown in Table \ref{tab: user study}, alignment rates consistently exceed 80\%, confirming strong agreement between metrics and human preferences. SQ aligns with MM-Dist and MDA, while NQ and FQ correspond to FID, M-Dist, Div, and GDA, ensuring its comprehensive evaluation.

\section{Conclusion}
In conclusion, we introduce CoheDancers, a novel framework designed for Group Music2Dance that systematically improves dance coherence by addressing synchronization, naturalness, and fluidity. This is achieved through the integration of the Cycle Consistency-based Dance Synchronization, Auto-Regressive Exposure Bias Correction, and Adversarial Training strategies. Additionally, we present I-Dancers, a comprehensive open-source dataset featuring dynamic dancer interactions, supported by dance kinesiology and music-dance retrieval methodologies. Our extensive experiments confirm the superior performance of CoheDancers and validate the robustness of I-Dancers. Looking ahead, future work could focus on integrating emotional expression and hand movements into dance generation. Additionally, developing personalized dance generation models that combine both music and text will be crucial for advancing this field.

% \bibliographystyle{named}
% \bibliography{ijcai24}

\begin{thebibliography}{}

\bibitem[\protect\citeauthoryear{Bengio \bgroup \em et al.\egroup }{2015}]{bengio2015scheduled}
Samy Bengio, Oriol Vinyals, Navdeep Jaitly, and Noam Shazeer.
\newblock Scheduled sampling for sequence prediction with recurrent neural networks.
\newblock {\em Advances in neural information processing systems}, 28, 2015.

\bibitem[\protect\citeauthoryear{Dhariwal \bgroup \em et al.\egroup }{2020}]{dhariwal2020jukebox}
Prafulla Dhariwal, Heewoo Jun, Christine Payne, Jong~Wook Kim, Alec Radford, and Ilya Sutskever.
\newblock Jukebox: A generative model for music.
\newblock {\em arXiv preprint arXiv:2005.00341}, 2020.

\bibitem[\protect\citeauthoryear{Foundation}{2024}]{blender2024}
Blender Foundation.
\newblock Blender 2.83, 2024.
\newblock Open Source 3D creation. Free to use for any purpose.

\bibitem[\protect\citeauthoryear{Gong \bgroup \em et al.\egroup }{2023}]{gong2023tm2d}
Kehong Gong, Dongze Lian, Heng Chang, Chuan Guo, Zihang Jiang, Xinxin Zuo, Michael~Bi Mi, and Xinchao Wang.
\newblock Tm2d: Bimodality driven 3d dance generation via music-text integration.
\newblock In {\em Proceedings of the IEEE/CVF International Conference on Computer Vision}, pages 9942--9952, 2023.

\bibitem[\protect\citeauthoryear{Guo \bgroup \em et al.\egroup }{2022}]{guo2022tm2t}
Chuan Guo, Xinxin Zuo, Sen Wang, and Li~Cheng.
\newblock Tm2t: Stochastic and tokenized modeling for the reciprocal generation of 3d human motions and texts.
\newblock In {\em European Conference on Computer Vision}, pages 580--597. Springer, 2022.

\bibitem[\protect\citeauthoryear{Kim \bgroup \em et al.\egroup }{2022}]{kim2022brand}
Jinwoo Kim, Heeseok Oh, Seongjean Kim, Hoseok Tong, and Sanghoon Lee.
\newblock A brand new dance partner: Music-conditioned pluralistic dancing controlled by multiple dance genres.
\newblock In {\em Proceedings of the IEEE/CVF Conference on Computer Vision and Pattern Recognition}, pages 3490--3500, 2022.

\bibitem[\protect\citeauthoryear{Le \bgroup \em et al.\egroup }{2023a}]{le2023controllable}
Nhat Le, Tuong Do, Khoa Do, Hien Nguyen, Erman Tjiputra, Quang~D Tran, and Anh Nguyen.
\newblock Controllable group choreography using contrastive diffusion.
\newblock {\em ACM Transactions on Graphics (TOG)}, 42(6):1--14, 2023.

\bibitem[\protect\citeauthoryear{Le \bgroup \em et al.\egroup }{2023b}]{le2023music}
Nhat Le, Thang Pham, Tuong Do, Erman Tjiputra, Quang~D Tran, and Anh Nguyen.
\newblock Music-driven group choreography.
\newblock In {\em Proceedings of the IEEE/CVF Conference on Computer Vision and Pattern Recognition}, pages 8673--8682, 2023.

\bibitem[\protect\citeauthoryear{Legrand and Ravn}{2009}]{legrand2009perceiving}
Doroth{\'e}e Legrand and Susanne Ravn.
\newblock Perceiving subjectivity in bodily movement: The case of dancers.
\newblock {\em Phenomenology and the Cognitive Sciences}, 8:389--408, 2009.

\bibitem[\protect\citeauthoryear{Li \bgroup \em et al.\egroup }{2021}]{li2021ai}
Ruilong Li, Shan Yang, David~A Ross, and Angjoo Kanazawa.
\newblock Ai choreographer: Music conditioned 3d dance generation with aist++.
\newblock In {\em Proceedings of the IEEE/CVF International Conference on Computer Vision}, pages 13401--13412, 2021.

\bibitem[\protect\citeauthoryear{Li \bgroup \em et al.\egroup }{2022}]{li2022danceformer}
Buyu Li, Yongchi Zhao, Shi Zhelun, and Lu~Sheng.
\newblock Danceformer: Music conditioned 3d dance generation with parametric motion transformer.
\newblock In {\em Proceedings of the AAAI Conference on Artificial Intelligence}, volume~36, pages 1272--1279, 2022.

\bibitem[\protect\citeauthoryear{Li \bgroup \em et al.\egroup }{2023}]{li2023finedance}
Ronghui Li, Junfan Zhao, Yachao Zhang, Mingyang Su, Zeping Ren, Han Zhang, Yansong Tang, and Xiu Li.
\newblock Finedance: A fine-grained choreography dataset for 3d full body dance generation.
\newblock In {\em Proceedings of the IEEE/CVF International Conference on Computer Vision}, pages 10234--10243, 2023.

\bibitem[\protect\citeauthoryear{Li \bgroup \em et al.\egroup }{2024a}]{li2024mvbench}
Kunchang Li, Yali Wang, Yinan He, Yizhuo Li, Yi~Wang, Yi~Liu, Zun Wang, Jilan Xu, Guo Chen, Ping Luo, et~al.
\newblock Mvbench: A comprehensive multi-modal video understanding benchmark.
\newblock In {\em Proceedings of the IEEE/CVF Conference on Computer Vision and Pattern Recognition}, pages 22195--22206, 2024.

\bibitem[\protect\citeauthoryear{Li \bgroup \em et al.\egroup }{2024b}]{li2024lodge}
Ronghui Li, YuXiang Zhang, Yachao Zhang, Hongwen Zhang, Jie Guo, Yan Zhang, Yebin Liu, and Xiu Li.
\newblock Lodge: A coarse to fine diffusion network for long dance generation guided by the characteristic dance primitives.
\newblock In {\em Proceedings of the IEEE/CVF Conference on Computer Vision and Pattern Recognition}, pages 1524--1534, 2024.

\bibitem[\protect\citeauthoryear{Liu \bgroup \em et al.\egroup }{2023}]{liu2023revisiting}
Ruyang Liu, Jingjia Huang, Ge~Li, Jiashi Feng, Xinglong Wu, and Thomas~H Li.
\newblock Revisiting temporal modeling for clip-based image-to-video knowledge transferring.
\newblock In {\em Proceedings of the IEEE/CVF Conference on Computer Vision and Pattern Recognition}, pages 6555--6564, 2023.

\bibitem[\protect\citeauthoryear{Loper \bgroup \em et al.\egroup }{2023}]{loper2023smpl}
Matthew Loper, Naureen Mahmood, Javier Romero, Gerard Pons-Moll, and Michael~J Black.
\newblock Smpl: A skinned multi-person linear model.
\newblock In {\em Seminal Graphics Papers: Pushing the Boundaries, Volume 2}, pages 851--866. 2023.

\bibitem[\protect\citeauthoryear{McFee \bgroup \em et al.\egroup }{2015}]{mcfee2015librosa}
Brian McFee, Colin Raffel, Dawen Liang, Daniel~PW Ellis, Matt McVicar, Eric Battenberg, and Oriol Nieto.
\newblock librosa: Audio and music signal analysis in python.
\newblock In {\em SciPy}, pages 18--24, 2015.

\bibitem[\protect\citeauthoryear{M{\"u}ller \bgroup \em et al.\egroup }{2005}]{muller2005efficient}
Meinard M{\"u}ller, Tido R{\"o}der, and Michael Clausen.
\newblock Efficient content-based retrieval of motion capture data.
\newblock In {\em ACM SIGGRAPH 2005 Papers}, pages 677--685. 2005.

\bibitem[\protect\citeauthoryear{Onuma \bgroup \em et al.\egroup }{2008}]{onuma2008fmdistance}
Kensuke Onuma, Christos Faloutsos, and Jessica~K Hodgins.
\newblock Fmdistance: A fast and effective distance function for motion capture data.
\newblock {\em Eurographics (Short Papers)}, 7, 2008.

\bibitem[\protect\citeauthoryear{Pavlakos \bgroup \em et al.\egroup }{2019}]{pavlakos2019expressive}
Georgios Pavlakos, Vasileios Choutas, Nima Ghorbani, Timo Bolkart, Ahmed~AA Osman, Dimitrios Tzionas, and Michael~J Black.
\newblock Expressive body capture: 3d hands, face, and body from a single image.
\newblock In {\em Proceedings of the IEEE/CVF conference on computer vision and pattern recognition}, pages 10975--10985, 2019.

\bibitem[\protect\citeauthoryear{Rallis \bgroup \em et al.\egroup }{2018}]{rallis2018embodied}
Ioannis Rallis, Apostolos Langis, Ioannis Georgoulas, Athanasios Voulodimos, Nikolaos Doulamis, and Anastasios Doulamis.
\newblock An embodied learning game using kinect and labanotation for analysis and visualization of dance kinesiology.
\newblock In {\em 2018 10th international conference on virtual worlds and games for serious applications (VS-Games)}, pages 1--8. IEEE, 2018.

\bibitem[\protect\citeauthoryear{Romero \bgroup \em et al.\egroup }{2022}]{romero2022embodied}
Javier Romero, Dimitrios Tzionas, and Michael~J Black.
\newblock Embodied hands: Modeling and capturing hands and bodies together.
\newblock {\em arXiv preprint arXiv:2201.02610}, 2022.

\bibitem[\protect\citeauthoryear{Siyao \bgroup \em et al.\egroup }{2022}]{siyao2022bailando}
Li~Siyao, Weijiang Yu, Tianpei Gu, Chunze Lin, Quan Wang, Chen Qian, Chen~Change Loy, and Ziwei Liu.
\newblock Bailando: 3d dance generation by actor-critic gpt with choreographic memory.
\newblock In {\em Proceedings of the IEEE/CVF Conference on Computer Vision and Pattern Recognition}, pages 11050--11059, 2022.

\bibitem[\protect\citeauthoryear{Siyao \bgroup \em et al.\egroup }{2023}]{siyao2023bailando++}
Li~Siyao, Weijiang Yu, Tianpei Gu, Chunze Lin, Quan Wang, Chen Qian, Chen~Change Loy, and Ziwei Liu.
\newblock Bailando++: 3d dance gpt with choreographic memory.
\newblock {\em IEEE Transactions on Pattern Analysis and Machine Intelligence}, 2023.

\bibitem[\protect\citeauthoryear{Siyao \bgroup \em et al.\egroup }{2024}]{siyao2024duolando}
Li~Siyao, Tianpei Gu, Zhitao Yang, Zhengyu Lin, Ziwei Liu, Henghui Ding, Lei Yang, and Chen~Change Loy.
\newblock Duolando: Follower gpt with off-policy reinforcement learning for dance accompaniment.
\newblock {\em arXiv preprint arXiv:2403.18811}, 2024.

\bibitem[\protect\citeauthoryear{Stathopoulos \bgroup \em et al.\egroup }{2024}]{stathopoulos2024score}
Anastasis Stathopoulos, Ligong Han, and Dimitris Metaxas.
\newblock Score-guided diffusion for 3d human recovery.
\newblock In {\em Proceedings of the IEEE/CVF Conference on Computer Vision and Pattern Recognition}, pages 906--915, 2024.

\bibitem[\protect\citeauthoryear{Sun \bgroup \em et al.\egroup }{2020}]{sun2020deepdance}
Guofei Sun, Yongkang Wong, Zhiyong Cheng, Mohan~S Kankanhalli, Weidong Geng, and Xiangdong Li.
\newblock Deepdance: music-to-dance motion choreography with adversarial learning.
\newblock {\em IEEE Transactions on Multimedia}, 23:497--509, 2020.

\bibitem[\protect\citeauthoryear{Tseng \bgroup \em et al.\egroup }{2023}]{tseng2023edge}
Jonathan Tseng, Rodrigo Castellon, and Karen Liu.
\newblock Edge: Editable dance generation from music.
\newblock In {\em Proceedings of the IEEE/CVF Conference on Computer Vision and Pattern Recognition}, pages 448--458, 2023.

\bibitem[\protect\citeauthoryear{Tsuchida \bgroup \em et al.\egroup }{2019}]{tsuchida2019aist}
Shuhei Tsuchida, Satoru Fukayama, Masahiro Hamasaki, and Masataka Goto.
\newblock Aist dance video database: Multi-genre, multi-dancer, and multi-camera database for dance information processing.
\newblock In {\em ISMIR}, volume~1, page~6, 2019.

\bibitem[\protect\citeauthoryear{Vaswani}{2017}]{vaswani2017attention}
A~Vaswani.
\newblock Attention is all you need.
\newblock {\em Advances in Neural Information Processing Systems}, 2017.

\bibitem[\protect\citeauthoryear{Wang \bgroup \em et al.\egroup }{2022}]{wang2022groupdancer}
Zixuan Wang, Jia Jia, Haozhe Wu, Junliang Xing, Jinghe Cai, Fanbo Meng, Guowen Chen, and Yanfeng Wang.
\newblock Groupdancer: Music to multi-people dance synthesis with style collaboration.
\newblock In {\em Proceedings of the 30th ACM International Conference on Multimedia}, pages 1138--1146, 2022.

\bibitem[\protect\citeauthoryear{Xu \bgroup \em et al.\egroup }{2023}]{xu2023actformer}
Liang Xu, Ziyang Song, Dongliang Wang, Jing Su, Zhicheng Fang, Chenjing Ding, Weihao Gan, Yichao Yan, Xin Jin, Xiaokang Yang, et~al.
\newblock Actformer: A gan-based transformer towards general action-conditioned 3d human motion generation.
\newblock In {\em Proceedings of the IEEE/CVF International Conference on Computer Vision}, pages 2228--2238, 2023.

\bibitem[\protect\citeauthoryear{Yang \bgroup \em et al.\egroup }{2024a}]{yang2024codancers}
Kaixing Yang, Xulong Tang, Ran Diao, Hongyan Liu, Jun He, and Zhaoxin Fan.
\newblock Codancers: Music-driven coherent group dance generation with choreographic unit.
\newblock In {\em Proceedings of the 2024 International Conference on Multimedia Retrieval}, pages 675--683, 2024.

\bibitem[\protect\citeauthoryear{Yang \bgroup \em et al.\egroup }{2024b}]{yang2024beatdance}
Kaixing Yang, Xukun Zhou, Xulong Tang, Ran Diao, Hongyan Liu, Jun He, and Zhaoxin Fan.
\newblock Beatdance: A beat-based model-agnostic contrastive learning framework for music-dance retrieval.
\newblock In {\em Proceedings of the 2024 International Conference on Multimedia Retrieval}, pages 11--19, 2024.

\bibitem[\protect\citeauthoryear{Yao \bgroup \em et al.\egroup }{2023}]{yao2023dance}
Siyue Yao, Mingjie Sun, Bingliang Li, Fengyu Yang, Junle Wang, and Ruimao Zhang.
\newblock Dance with you: The diversity controllable dancer generation via diffusion models.
\newblock In {\em Proceedings of the 31st ACM International Conference on Multimedia}, pages 8504--8514, 2023.

\bibitem[\protect\citeauthoryear{Zhang \bgroup \em et al.\egroup }{2024}]{zhang2024bidirectional}
Canyu Zhang, Youbao Tang, Ning Zhang, Ruei-Sung Lin, Mei Han, Jing Xiao, and Song Wang.
\newblock Bidirectional autoregessive diffusion model for dance generation.
\newblock In {\em Proceedings of the IEEE/CVF Conference on Computer Vision and Pattern Recognition}, pages 687--696, 2024.

\bibitem[\protect\citeauthoryear{Zhou \bgroup \em et al.\egroup }{2019}]{zhou2019continuity}
Yi~Zhou, Connelly Barnes, Jingwan Lu, Jimei Yang, and Hao Li.
\newblock On the continuity of rotation representations in neural networks.
\newblock In {\em Proceedings of the IEEE/CVF conference on computer vision and pattern recognition}, pages 5745--5753, 2019.

\bibitem[\protect\citeauthoryear{Zhuang \bgroup \em et al.\egroup }{2022}]{zhuang2022music2dance}
Wenlin Zhuang, Congyi Wang, Jinxiang Chai, Yangang Wang, Ming Shao, and Siyu Xia.
\newblock Music2dance: Dancenet for music-driven dance generation.
\newblock {\em ACM Transactions on Multimedia Computing, Communications, and Applications (TOMM)}, 18(2):1--21, 2022.

\bibitem[\protect\citeauthoryear{Zhuang \bgroup \em et al.\egroup }{2023}]{zhuang2023gtn}
Haolin Zhuang, Shun Lei, Long Xiao, Weiqin Li, Liyang Chen, Sicheng Yang, Zhiyong Wu, Shiyin Kang, and Helen Meng.
\newblock Gtn-bailando: Genre consistent long-term 3d dance generation based on pre-trained genre token network.
\newblock In {\em ICASSP 2023-2023 IEEE International Conference on Acoustics, Speech and Signal Processing (ICASSP)}, pages 1--5. IEEE, 2023.

\end{thebibliography}

\end{document}